# Digital Radar for Collision Avoidance and Automatic Cruise Control in Transportation

Rabindranath Bera, Sourav Dhar, Debdatta Kandar

Sikkim Manipal Institte of Technology, Sikkim Manipal University, majitar, rangpo, East Sikkim 737132
E-mail: dhar.sourav@gmail.com

**Abstract:**
*A proper remote sensing device is required for automatic cruise control (ACC) to avoid collision in transportation system. In this paper we proposed a direct sequence spread spectrum (DSSS) radar for remote sensing in intelligent transporation system(ITS). We have successfully detected single target and through 1D radar imaging we are capable to separate multiple targets. We have also implemented DSSS radar using software defined radio (SDR) and successfully detected a single target.*

## I. INTRODUCTION

**V**EHICLE and highway automation is believed to reduce the risk of accidents, improve safety, increase capacity, reduce fuel consumption and enhance overall comfort and performance for drivers. It is believed that more automated automobiles relieve the driver from many undesirable routines of driving task. It has also been known that many of the car accidents are due to human errors [1] . Therefore, the conclusion has been that with a robust automated system the chance of car accidents can be reduced. With the overwhelming increase in the number of vehicles on the road another concern has been road capacity. Some kind of automation that would help to safely increase traffic flow has been considered as one potential solution to congested highways. A smoother cruise with an automated system can reduce fuel consumption and engine wear.

The main objective of automated vehicle system which is known as intelligent transportation system (ITS) was to improve highway capacity and safety with automation in highway and vehicle level [2].

In this work we are focusing the detection and imaging of nearby vehicle. Target detection and sensing is one of the important aspect for ITS. There are several remote sensing techniques in use including radar for target detection in ITS. There are various kinds of sensors available for ITS applications. RADAR is having the advantage of high detection range, high range resolution, lower algorithmic complexity with moderate hardware cost. Also, it actively works in darkness, rainy and foggy conditions with high accuracy [3].

The authors would like to highlight the use of Direct Sequence Spread Spectrum (DSSS) based digital radar in ITS. The basic DSSS radar for target detection and Imaging is developed at 2.4 GHz carrier and is now operational at Sikkim Manipal Institute of Technology (SMIT), INDIA. Several experiments are successfully conducted at the rooftop of SMIT using the radar for the detection and characterization of targets. Three levels of hardware experiments are conducted. Level 1: Detection of Single target like flat plats, persons, Foiled Globes, banana. Level 2: Radio contour mapping of the area in front of the radar. Level 3: Detection of multiple targets.

Outcome of the hardware experiment is quite interesting. It is capable of detection of multiple targets but unable to separate them. To overcome this limitation, MATLAB/ SIMULINK based simulation is performed. Instead of single frequency carrier of 2.4 GHz, the step frequency mode will be the objective for target separation and imaging [4]. We have successfully developed this end to end radar system simulation. In the 1$^{st}$ phase, attempt is made on simulation of single carrier frequency (2.4 GHz ) using this target model to detect a single target. Extraction of phase change and signal attenuation made by the target is achieved with very good accuracy. After successful detection of single target, the simulation is extended to find the effect of multiple targets. Alike hardware simulation additive and subtractive effect for resultant attenuation is observed here too.

From the above discussion we find that single frequency radar is unable to separate the different targets.

In the 2$^{nd}$ phase of attempt, Frequency stepping is used to separate the targets and their imaging. Thus a block for frequency stepping is added to the RF block of transmitter and receiver of the main simulation model. Frequency stepping method is simulated for proper detection and separation of different targets at very preliminary level. More elaborate works yet to be done for proper imaging of targets using simulation.

The 3$^{rd}$ phase of radar signal processing simulation is also performed using rake processing at the receiver. All the radar signal processing discussed above cannot be implemented using additional hardware or software in hardware simulation model. A need for SDR ( Software Defined Radio) is thus justified to be used as DSP tool and all DSP processing can be realized using it. Most of the DSP simulations are partially implemented using SDR. In essence, the development of the digital radar is almost ready for its deployment at the vehicle for collision avoidance and ACC ( Automatic Cruise Control) [5].

## II. HARDWARE EXPERIMENT OF DIGITAL RADAR

The basic DSSS radar for target detection, Imaging and RCS measurement, radar is developed and is now operational at SMIT. Several experiments are successfully conducted at the rooftop of SMIT using the radar for the detection and characterization of targets. In order to measure the target parameters we graduated roof top floor into two dimensional co-ordinates. The square plates at the roof top having area of 1sqft each. This helped us in Radio mapping. The foiled globe is taken as a target because of it's spherical nature. The target return in form of received signal strength (SNR) is noted with the target placed at different co-ordinates. The background signal strength without any target is noted to be -78 dbm. Fig. 1. depicts the graphical presentation of the radio mapping. The blue line zone has signal strength of -70 dbm to -75dbm. The green line zone shows signal level of -60 dbm to -69dbm. There is a strongest zone at which the signal strength is measured to be -55dbm. A contour map is plotted covering an area of (20 feet x 25 feet) in front of radar. Target imaging requires amplitude, phase and frequency information of individual scatterer. Radio contour mapping will help us radar imaging in future.

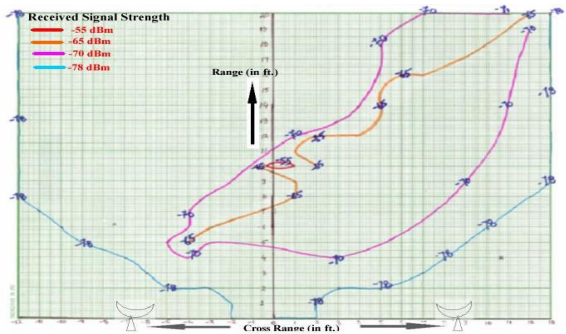

Fig.1: Graphical representation of Radar Mapping.

The signal strength varies with the number of scatterers. The variation of signal strength with different number of scatterers is shown in the Table I. It is interesting to note that the received signal strength is additive or subtractive depending on the positions of the Scatterers.

The additive or subtractive nature of signal strength is due to multipath signals. Multipath signals are added at different phase which, in effect, is reducing the main signal. Fig. 2 shows the multipath effect using two targets. Effectively, instead of multiple targets resolution, they may be treated as multi path targets and presence of one target will influence to the other in additive and subtractive way. The radar is not able to resolve the effect of multiple target. *This is, therefore, the limitation of DSSS radar hardware simulation.* Hence the Radar needs to be modified in term of better signal processing capabilities using Adaptive Equalizer and Rake processing.

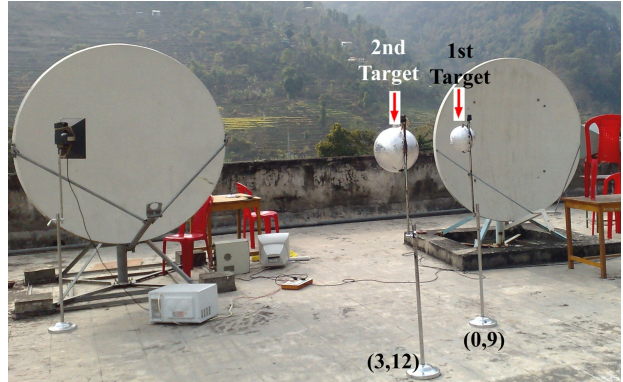

Fig.2: Multipath effects using 2 targets

**Table I: Variation in Signal Strength for multiple Targets**

| Coordinates of Desired Target (ft) | Signal Strength of single object (dbm) | Coordinates of second object (ft) | Combined Signal Strength (dBm) | Remarks |
|---|---|---|---|---|
| 0,9 | -63 | 3,12 | -58 | Additive |
| -3,6 | -59 | 3,12 | -57 | Additive |
| -3,6 | -59 | 0,9 | -64 | Subtractive |
| 0,9 | -63 | -3,6 | -54 | Additive |

## III. SOFTWARE SIMULATION OF DIGITAL RADAR

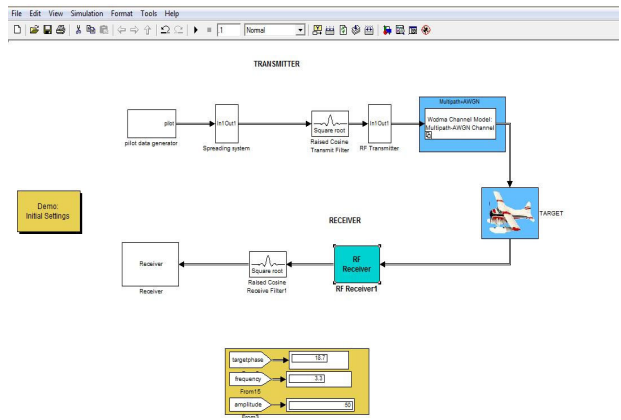

**Fig. 3:** Simulation Model of DSSS radar used for target detection.

The simple way to implement a target model is to represent it as a collection of point scatterers. Each scatterer, therefore, will be characterized by a distance from radar and the strength of reflection (and thus, a path loss and a delay associated with that distance). In the DSSS radar (Fig 3), the target is modeled as a cascade of path loss and phase change as shown in Fig. 4. In this target model a provision for doppler frequency shift is also included which is mainly due to the motion of the target. Since in the open orange measurement, the target will be static, the Doppler frequency is kept as 0.1 Hz for all target detections.

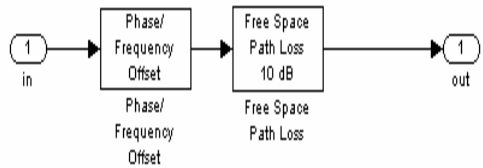

Fig . 4: Target model in DSSS radar simulation

Simulation is conducted at 2.4 GHz carrier frequency using this target model to detect a single target on the basis of signal attenuation and phase change made by the target. Table II is showing some data obtained during the simulation. For performing simulation it is assumed that a large target will attenuate the signal less and will make less phase change compared to a smaller target. Extraction of phase change and signal attenuation made by the target is achieved with very good accuracy. It can also be observed that the accuracy of measurement is better for bigger target compared to smaller target.

**Table II: Results obtained for single target detection**
*Transmission frequency= 2.4 GHz*
*Doppler Frequency offset =0.1 Hz*

| S L . N o . | Single Target | | | |
|---|---|---|---|---|
| | Phase Change introduced by the Target (degree) | Measured Phase Change at the receiver (degree) | Signal Attenuation Introduced by the target (dB) | Attenuation measured From Received Signal(dB) |
| 1. | 1 | 1.167 | 2 | 2.09 |
| 2. | 10 | 10.1 | 4 | 4.026 |
| 3. | 15 | 15.06 | 6 | 5.986 |
| 4. | 18 | 18.05 | 7 | 6.972 |
| 5. | 20 | 20.04 | 8 | 7.961 |
| 6. | 25 | 25.03 | 9 | 8.953 |
| 7. | 30 | 30.02 | 10 | 9.946 |
| 8. | 40 | 40.02 | 11 | 10.94 |
| 9. | 45 | 45.01 | 12 | 11.94 |

### IV. MULTIPLE TARGET DETECTION

After successful detection of single target, the simulation is extended to find the effect of multiple targets. Thus an additional similar target model block is required to be inserted in the original simulation as in Fig.1.4. Here two targets are placed in front of the radar. One target kept static and other has been kept at different places. The resultant signal attenuation is measured. Since the simulation is performed at 2.4 GHz having λ= 12.5 cms. Since λ corresponds to $360^o$ phase shift, thus it can be said that a relative distance of 12.5 cms corresponds $360^o$ phase shift. This logic is used to find the resultant signal attenuation at different relative distances between two targets. In this simulation both targets can make 10 dB attenuation to the transmitted signal individually. One of them kept at a fixed place and position of the other is varied [ Fig. 5]. Effectively this will give different relative distance (d1, d2, d3 etc.) and hence phase shift will vary. Two scattered signals are interfering with each other at different phases and thus the resultant signal attenuation will be different at different relative distances. Here relative distance is varied from 0 to 3 meters and resultant signal attenuations are noted, where relative distance brings both the targets in phase there resultant attenuation noted is minimum (4.006 dB). On the other hand, for $180^0$ out of phase maximum signal attenuation (161.3 dB) is observed. Since our desired target attenuation is 10 dB hence it can be said that the effect of multiple target is additive for resultant attenuation less than 10dB and subtractive for resultant attenuation more than 10dB [similar observations are noted in hardware experiment as in Table I].

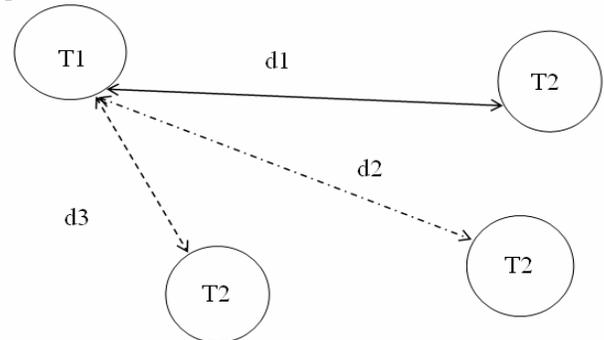

Fig. 5: Target 1 is fixed and Target 2 is varied.

For example, let us assume the transmitted signal strength is +5dB. For a single target which produces 10dB attenuation, received signal strength will be -5dB. Now for two targets of same kind if resultant attenuation is 4dB then received signal strength will be +1dB. This is an example of additive multiple target effect. Similarly, if resultant attenuation is 15dB then received signal strength will be -10dB which is a subtractive effect.

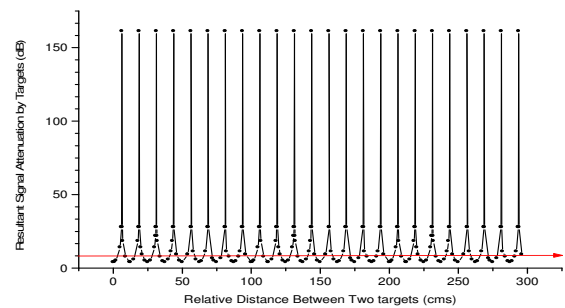

Fig. .6: Variation of resultant signal attenuation with relative distance.

In Fig. 6, points bellow the red line is representing the additive zone and above red line is subtractive zone for our simulation with the target of 10dB attenuation. Some simulated data are shown in Table III.

**Table III: Results obtained from Software Simulation of DSSS radar for multiple target detection**
*Two Targets*
*Transmission frequency= 2.4 GHz*
*Signal Attenuation by the target I= 10dB*
*Signal Attenuation by the target II= 10dB*
*Phase Change Made by the Target I= 0 degree (Static)*
*Doppler Frequency offset =0.1 Hz*

| Sl. No. | Relative Distance between target I and Targer II (cms) | Attenuation measured from Received Signal(dB) | Remarks |
|---|---|---|---|
| | 0 | 4.006 | Additive |
| | 20 | 14.11 | Subtractive |
| | 40 | 5.809 | Additive |
| | 134 | 7.853 | Additive |
| | 172 | 6.701 | Additive |
| | 244 | 27.94 | Subtractive |

## V. RELATION OF MATLAB SIMULATION AND HARDWARE EXPERIMENT OF DSSS RADAR

Table I is showing the results obtained from the hardware experiment of DSSS radar performed at SMIT roof top. It is clear from these results that multiple target is producing some additive effect at some distances and subtractive effects at some other distances. Our simulation result is supporting the hardware experiment. Referring Table I, one target at (-3,6) coordinate and other at (3,12) producing an additive effect. Each coordinate corresponds to 1ft distance. Thus the relative distance between target 1 and target 2 for Sl.No. 2 of Table I, can be calculated as,

$R = [(x_1-x_2)^2 + (y_1-y_2)^2]^{1/2} = [(-3-3)^2 + (6-12)^2]^{1/2} = [36+36]^{1/2} = 8.4852$ ft = 258.6288 cms

From the plot shown in Fig. 6, we can observe that multi target effect is additive at 258.6288 cms relative distance. Similarly, the relative distance between target 1 and target 2 for Sl.No. 3 can be calculated as,

$R = [(x_1-x_2)^2 + (y_1-y_2)^2]^{1/2} = [(-3-0)^2 + (6-9)^2]^{1/2} = [9+9]^{1/2} = 4.2426$ ft = 129.3144 cms

From the plot shown in Fig. 6, we can observe that multi target effect is subtractive at 129.3144 cms relative distance. Hence it is clear that the MATLAB simulation and Hardware Experiment of DSSS radar are producing similar results.

## VI. SEPARATION OF TARGETS

From the above discussion we find that this simulation model is capable of detecting the effect of multiple targets but unable to separate the target. Frequency stepping is a method that is used to separate the targets and their imaging like [4]. Thus a block for frequency hopping is added to the RF block of transmitter and receiver of Fig. 3. The modified RF block is shown in Fig. 7. The frequency hopping spectrum is shown Fig. 8. The frequency is stepped over 3GHz bandwidth.

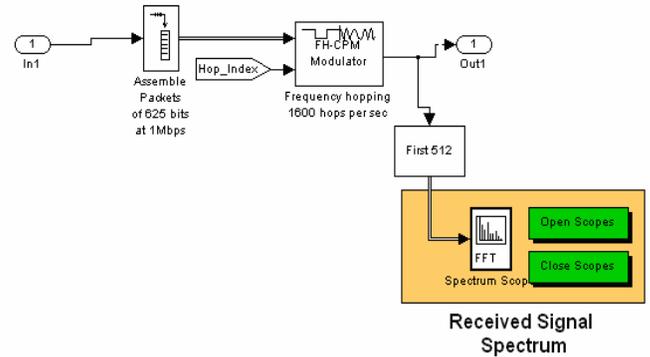
Fig.7 details of frequency sweeping.

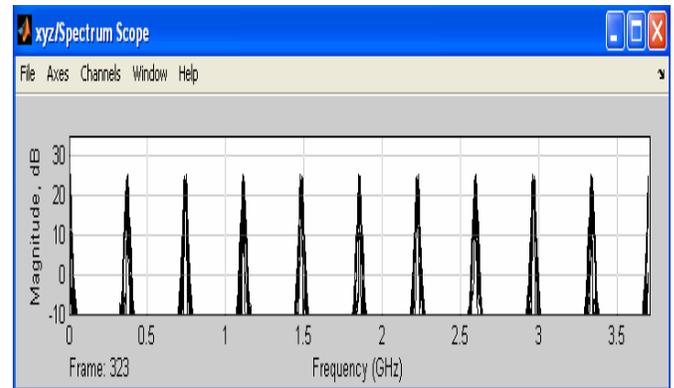
Fig.8 Frequency Hopped signals.

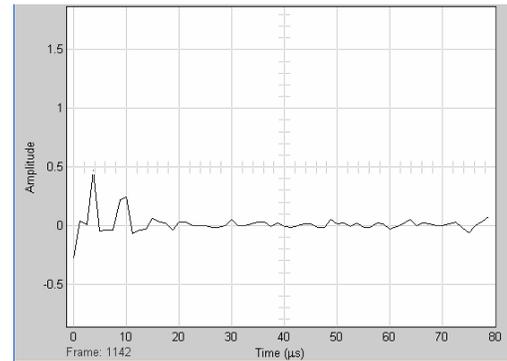
Fig. 9: 1D imaging through software

1D radar imaging (Fig. 9) is successfully done using this simulation model. It evident from Fig. 9 that the two targets are now separable.

## VII. SDR IMPLEMENTATION

Implementation of DSSS radar is done using Lyrtech SignalWave software defined radio (SDR) kit. This SDR board consists of a DSP section and an FPGA section. DSP and FPGA are programmed separately. SIMULINK is used for Model based design to programme the SDR. Simulation model of Fig. 3 has been modified a bit for implementation through SDR and shown in Fig. 10. Loop back test is done and transmitted and received power spectrum are observed for different kind of signal attenuation introduced by the target (Fig. 11 and Fig. 12).

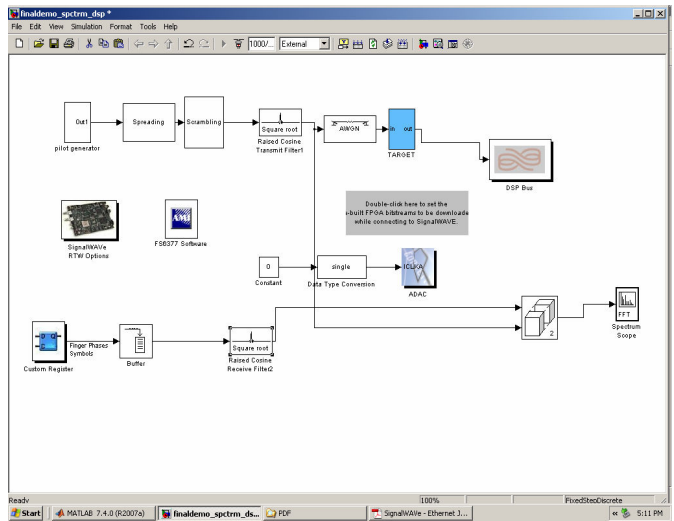
**Fig. 10:** DSP Section for the DSSS radar system

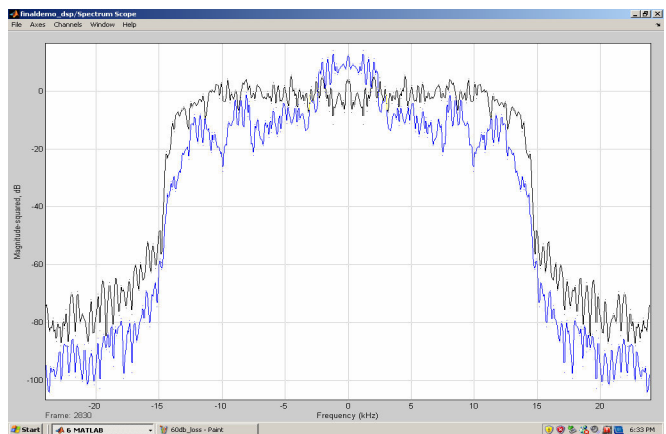
Fig.11. Spectrums of Tx and Rx signal power through SDR (0dB path loss)

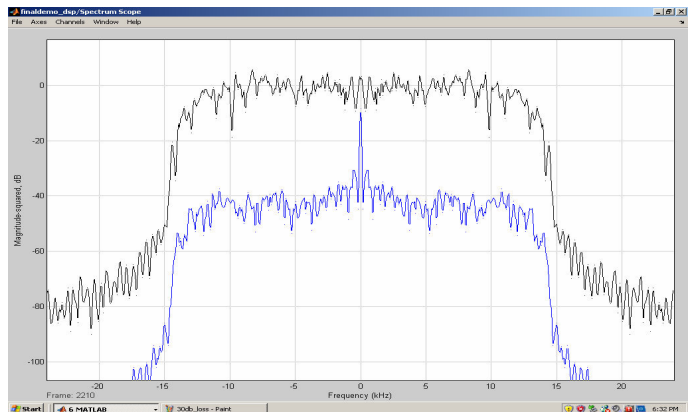
Fig.12. Spectrums of Tx and Rx signal power through SDR (30dB path loss)

In Fig.11 and Fig.12, black graph stands for transmitted signal power and blue stands for received signal power. Thus the detection of targets are implemented though SDR successfully.

## VI. CONCLUSION

In this work we have successfully detected single target using hardware as well as in simulation. 1D imaging is successfully done using frequency stepping method. Thus separation of multiple targets is possible using radar imaging. 3D imaging will be able provide all the information about the target. Digital radar implementation is done for single target detection using SDR. But, here we could only extract the signal attenuation. Phase and frequency offset extraction are yet to be implemented. DSSS radar using SDR will be useful for implementing it in a car. This implementation will be able to provide information (distance and type or size of the vehicle) about the nearby vehicle for automatic cruise control (ACC).

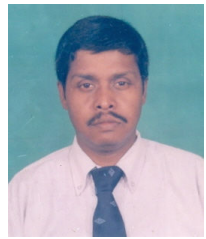

**Rabindra Nath Bera**: Born in 1958 at Kolaghat , West Bengal, INDIA. Received his B. Tech, M. Tech & Ph.D (Tech) from the Institute of Radiophysics & Electronics, The University of Calcutta, in the year 1982,1985 & 1997 respectively. Currently working as Dean (R&D) and Head of the Deparment, Electronics & Communication Engineering, Sikkim Manipal University,Sikkim. Microwave/ Millimeter wave based Broadband Wireless Mobile Communication and Remote Sensing are the area of specializations.

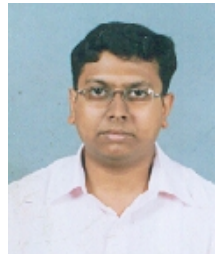

**Sourav Dhar:** Born in 1980 at Raiganj, West Bengal, INDIA. Received B.E from Bangalore institute of Technology and M.Tech from Sikkim Manipal Institute of Technology in the year 2002 and 2005 respectively. Currently working as a Senior Lecturer , Department of E&C Engineering , Sikkim Manipal  University, Sikkim, India. Broadband Wireless Mobile Communication and Remote Sensing are the area of specializations.